\documentstyle[graphicx, 12pt]{article}
\begin{document}

\small
\hoffset=-1truecm
\voffset=-2truecm
\title{\bf The greybody factor for scalar fields
in the Schwarzschild spacetime with an $f(R)$ global monopole}
\author{Jingyun Man \hspace {1cm} Hongbo Cheng\footnote
{E-mail address: hbcheng@ecust.edu.cn}\\
Department of Physics, East China University of Science and
Technology,\\ Shanghai 200237, China\\
The Shanghai Key Laboratory of Astrophysics, Shanghai 200234,
China}

\date{}
\maketitle

\begin{abstract}
The greybody factor of massless scalar fields in the
four-dimensional Schwarzschild spacetime involving an $f(R)$
global monopole is derived. We show how the monopole parameter and
the deviation from the standard general relativity adjust the
greybody factor. We also demonstrate that the effects from the
global monopole and $f(R)$ gravity theory are manifest in the
energy emission rate and the generalized absorption cross section
of the scalar fields.
\end{abstract}

\vspace{7cm} \hspace{1cm} PACS number(s): 04.70.Bw, 04.70.Dy,
14.80.Hv, 11.80.-m

\newpage

\noindent \textbf{I.\hspace{0.4cm}Introduction}

In any gravitational theory black holes are the ones of the most
relevant objects and have been the focuses while a lot of
contributions were paid to them. The black holes could exist at
the centre of galaxies [1]. In particle collide the black holes
could generate [2, 3]. There are different directions to explore
the black holes. The black holes thermodynamics such as their
entropy [4-6], thermal radiation [7], phase transition [8-13] were
investigated. In addition, a lot of efforts were contributed to
the scattering and absorption properties of waves in the spacetime
of black hole belonging to the asymptotically flat case [14-20].
The greybody factor defined as the probability for a given wave
coming in from infinity to be absorbed by the black hole is
directly connected to the absorption cross section and is also
discussed [21, 22]. The greybody factors in the four-dimensional
Schwarzschild-de Sitter spacetime were computed in the case of
minimally coupled scalar field [22-26] or nonminimally coupled
ones [27] respectively. In the Schwarzschild-de Sitter spacetime
the greybody factor with vanishing angular quantum number in the
infrared limit tends to a positive constant for a minimally
coupled massless scalar field and goes to zero for the case of a
nonminimally coupled ones.

During the process of the vacuum phase transition in the early
universe, various kinds of topological defects like domain walls,
cosmic strings and monopoles were generated from the breakdown of
local or global gauge symmetries [28, 29]. Among these topological
defects, a global monopole as a spherical symmetric topological
defect occurred in the process of phase transition of a system
consisting of a self-coupling triplet of a scalar field whose
original global $O(3)$ symmetry is spontaneously broken to $U(1)$.
It was found that the metric outside a monopole has a deficit
solid angle [30]. Buchdahl put forward a modified gravity theory
named as $f(R)$ gravity to explain the accelerated expansion of
the universe instead of adding unknown form of dark energy or dark
matter [31-34]. The metric around a gravitational source involving
a global monopole within the frame of $f(R)$ gravity theory has
been studied [35]. The classical motion of a massive test particle
around the gravitational object with an $f(R)$ global monopole is
discussed [36]. We examine the gravitational lensing for the same
object in the strong field limit [37]. We also investigate the
thermodynamic quantities of this kind of black hole [38].

The purpose of this paper is to compute the greybody factor for
massless scalar fields in the environment of a static and
spherically symmetric black hole swallowing an $f(R)$ global
monopole. This kind of gravitational sources could contain the
global monopole while survive in the universe with accelerated
expansion described with the help of $f(R)$ theory. The features
of global monopole and $f(R)$ issue may appear simutaneously. We
should take into account the role played by the geometrical
features of both the global monopole and the modified gravity
theory. We can explore the global monopole by means of the
greybody factor also corrected by the $f(R)$ gravity theory. It is
also significant that we study the greybody factors for massless
scalar fields propagating outside this kind of the black hole to
understand the $f(R)$ theory in a new direction. We wish to show
how the $f(R)$ theory modifies the factors. The influence from the
modified gravity on the energy emission and absorption cross
section will also be shown. This could be a new window to observe
the effects from global monopole and $f(R)$ theory. At first we
introduce the metric outside a black hole containing a global
monopole in the context of $f(R)$ gravity theory. A massless
scalar field for general coupling $\xi$, propagating in this
spacetime induced by global monopole and the deviation from
general relativity except for the source mass is considered. The
corresponding greybody factors are also scrutinized. Having
derived a complementary low-frequency approximation of the
greybody factors with arbitrary coupling $\xi$, we discuss how the
deviation of general relativity relates to the factors
analytically and numerically. Further the influence from $f(R)$
issue on the energy emission and the generalized absorption cross
section will be presented. Finally the discussions and conclusions
are listed.

\vspace{0.8cm} \noindent \textbf{II.\hspace{0.4cm}The equation of
massless scalar field with coupling $\xi$ around the massive
source with an $f(R)$ global monopole}

According to Ref. [35], the metric describing the background of
the black hole involving an $f(R)$ global monopole is,

\begin{equation}
ds^{2}=f(r)dt^{2}
-\frac{1}{f(r)}dr^{2}
-r^{2}(d\theta^{2}+\sin^{2}\theta d\varphi^{2})
\end{equation}

\noindent with

\begin{equation}
f(r)=1-8\pi G\eta^{2}-\frac{2GM}{r}-\psi_{0}r
\end{equation}
\noindent where $M$ is the mass and $G$ is the Newton constant.
Here the Lagrangian for the global monopole is
$\mathcal{L}=\frac{1}{2}(\partial_{\mu}\phi^{a})(\partial^{\mu}\phi^{a})
-\frac{1}{4}\lambda(\phi^{a}\phi^{a}-\eta^{2})^{2}$ with
parameters $\lambda$, $\eta$ and the ansatz for the triplet of
field configuration $\phi^{a}=\eta h(r)\frac{x^{a}}{r}$ while
$x^{a}x^{a}=r^{2}$ and $a=1, 2, 3$. $h(r)$ is a dimensionless
function to be determined by its equation of motion [30]. This
model has a global $O(3)$ symmetry, which is spontaneously broken
to $U(1)$. Here subject to Ref. [35, 36] $f(R)$ is an analytical
function of Ricci scalar $R$ and satisfies
$\frac{df(R)}{dR}=1+\psi_{0}r$. Now the correction to the general
relativity is limited as $\psi_{0}r\ll1$. The tiny factor
$\psi_{0}$ reflects the deviation of standard general relativity
and then the $f(R)$ gravity model could explain the cosmic
acceleration. It should be pointed out that the model parameter
$\eta$ is of the order $10^{16}GeV$ for a typical grand unified
theory, which means $8\pi G\eta^{2}\approx10^{-5}$. If we choose
$\psi_{0}=0$ excluding the modification from $f(R)$ theory, the
metric (2) will recover to be the result by Barriola and Vilenkin
[30].

We find that the black hole horizion and the cosmological horizon
of metric (2) are located at,

\begin{equation}
r_{\pm}=\frac{(1-8\pi G\eta^{2})\pm\sqrt{(1-8\pi
G\eta^{2})^{2}-8\psi_{0}GM}}{2\psi_{0}}
\end{equation}

\noindent which reduce to a black hole with a global monopole who
has only one event horizon $r_{H}=\frac{2GM}{1-8\pi G \eta ^{2}}$.
In this metric (2) we study a massless scalar field
$\Phi(x^{\mu})$ coupled to the gravitational field. The action for
this scalar field is [40],

\begin{equation}
S=\frac{1}{2}\int d^{4}x\sqrt{-g}(\nabla_{\mu}\Phi\nabla^{\mu}\Phi
+\xi R\Phi^{2})
\end{equation}

\noindent where $g$ is the determinant of the spacetime metric and
$\xi$ is the coupling between the scalar field and the
gravitational fields. Equivalently this massless scalar field
satisfies the Klein-Gordon equation,

\begin{equation}
(\nabla_{\mu}\nabla^{\mu}+\xi R)\Phi(x)=0
\end{equation}

\noindent In order to solve the Eq. (5), we write the solution in
the following form [40],

\begin{eqnarray}
\Phi(x)=\Phi(t, r, \theta, \varphi)\hspace{0.5cm}\nonumber\\
=\frac{\psi_{\omega l}(r)}{r}Y_{lm}(\theta, \varphi)e^{-i\omega t}
\end{eqnarray}

\noindent where $\omega$ is the frequency. $Y_{lm}(\theta,
\varphi)$ are the scalar spherical harmonics. The radial part of
the solution to Eq. (5) obeys the following equation,

\begin{equation}
f\frac{d}{dr}(f\frac{d\psi_{\omega l}}{dr})-V(r)\psi_{\omega
l}+\omega^{2}\psi_{\omega l}=0
\end{equation}

\noindent where

\begin{equation}
V(r)=f[\xi f''
+(4\xi+1)\frac{f'}{r}+2\xi\frac{f}{r^{2}}
+\frac{l(l+1)-2\xi}{r^{2}}]
\end{equation}

\noindent is the potential and the prime stands for the derivative
with respect to $r$ and $l$ is the angular quantum number. The
terms with coupling constant in Eq. (7) changes the evolution of a
scalar field in the background of a Schwarzschild black hole with
a $f(R)$ global monopole.

\vspace{0.8cm} \noindent \textbf{III.\hspace{0.4cm}The greybody
factor for massless scalar field with coupling $\xi$ around the
massive source with an $f(R)$ global monopole}

We plan to determine analytically the greybody factors for the
radiation of a emitted scalar particle generated by a
Schwarzschild black hole with a $f(R)$ global monopole. We follow
the procedure of Ref. [27, 39] to investigate the metric. The
nonlinear equation (7) is very difficult to obtain the analytic
solution directly. We adopt the low-frequency approximation by
solving the radial equation in near-region where $r$ is close to
the black hole horizon and far-region close to the cosmological
horizon and matching two results in the intermediate region. Both
the energy emission rate and the absorption cross section are
dependent on the ratio of the coefficient $A_{\omega l}^{out}$ and
$A_{\omega l}^{in}$ yielded in solutions of the nonlinear
equation. The condition that two wave solutions in near and far
region can be overlapped in a intermediate region is $\omega \ll
1/MG$, which demands low energy for the emission of a scalar field
in black hole spacetime. We define the near-region where
$r-r_{-}\ll\frac{1}{\omega}$, and the far-region as $r-r_{-}\ll
MG$. Then one can match two kinds of wave in the overlapping range
$MG\ll r-r_{-}\ll \frac{1}{\omega}$.

The metric $f(r)=0$ when the radial coordinate goes to either the
inner horizon or the outer horizon, which leads a vanished
potential. Near the inner event horizon of the black hole with an
$f(R)$ global monopole, for no outgoing flux across the black hole
horizon, the solution to Eq. (7) can be written as [27],

\begin{equation}
\psi_{\omega l}(r)\approx A^{tr}_{\omega l}e^{-i\omega r_{\ast}}
\end{equation}

\noindent where $r_{\ast}$ is the tortoise coordinate defined by
$\frac{dr_{\ast}}{dr}\equiv\frac{1}{f}$. The solution near the
outer event horizon of this kind of black hole can be denoted as,

\begin{equation}
\psi_{\omega l}(r)\approx A_{\omega l}^{in}e^{-i\omega r_{\ast}}
+A_{\omega l}^{out}e^{i\omega r_{\ast}}
\end{equation}

\noindent where $A_{\omega l}^{in}$ and $A_{\omega l}^{out}$
represent the amplitudes of the incoming wave and the outgoing
wave respectively. According to the definition, the greybody
factors can be written as [27],

\begin{eqnarray}
\gamma_{l}(\omega)=|\frac{A_{\omega l}^{tr}}{A_{\omega
l}^{in}}|^{2}
=1-|\frac{A_{\omega l}^{out}}{A_{\omega l}^{in}}|^{2}
\end{eqnarray}

\noindent because of $|A_{\omega l}^{in}|^{2}=|A_{\omega
l}^{tr}|^{2}+|A_{\omega l}^{out}|^{2}$ from flux conservation. If
we estimate the coefficients $A_{\omega l}^{in}$ and $A_{\omega
l}^{out}$, the greybody will be confirmed.

A black hole act like a greybody not a perfect blackbody because
the spectrum of a particle emitted by a black hole or a blackbody
are significantly different in low-frequency while in
high-frequency they agree well. Analytically the absorption
probability with vanished angular quantum number and no coupling
to gravitational field are given by
$\frac{4r_{+}^{2}r_{-}^{2}}{(r_{+}^{2}+r_{-}^{2})^{2}}$ [39].
However, the nonminimal coupling constant contributes a zero
greybody factor with $l=0$ when $\omega\to 0$ [27]. For these
reasons, absorption probability in low-energy, $\omega \ll 1/MG$,
contains the information about the structure of spacetime around a
black hole.

It is important to research on the amplitudes of the incoming wave
and the outgoing wave. Let

\begin{equation}
R_{\omega l}(r)=\frac{\psi_{\omega l}(r)}{r}
\end{equation}

\noindent then rewrite the Eq. (7) as,

\begin{equation}
\frac{d}{dr}(r^{2}f\frac{dR_{\omega l}}{dr})+ [-\xi
Rr^{2}+\frac{\omega^{2}}{f}r^{2}-l(l+1)]R_{\omega l}=0
\end{equation}

\noindent where the Ricci scalar curvature is

\begin{equation}
R=f''+\frac{4f'}{r}+\frac{2f}{r^{2}}-\frac{2}{r^{2}}
\end{equation}

\noindent We solve the equation of radial parts like Eq. (13) near
the event horizons because the amplitudes exist in these regions.
When the radial coordinate $r\longrightarrow r_{-}$, the $f(R)$
gravity term $\psi_{0}r$ in metric verges to zero, then Eq. (13)
will be approximated to be,

\begin{equation}
y(1-y)\frac{d^{2}R_{\omega l}^{n}}{dy^{2}}+(1-y)\frac{dR_{\omega
l}^{n}}{dy}+[\frac{16\pi\xi G\eta^{2}-l(l+1)}{1-8\pi G\eta^{2}}
\frac{1}{1-y}+\frac{\omega^{2}r_{-}^{4}}{(2GM)^{2}}
\frac{1-y}{y}]R_{\omega l}^{n}=0
\end{equation}

\noindent where

\begin{equation}
y=1-\frac{2GM}{1-8\pi G\eta^{2}}\frac{1}{r}
\end{equation}

\noindent Here $R_{\omega l}(r)$ near the inner horizon is written
as $R_{\omega l}^{n}(r)$. We choose,

\begin{equation}
R_{\omega l}^{n}=y^{i\varpi}(1-y)^{L+1}F_{\omega l}
\end{equation}

\begin{equation}
L(L+1)=\frac{l(l+1)-16\pi\xi G\eta^{2}}{1-8\pi G\eta^{2}}
\end{equation}

\begin{equation}
\varpi=\frac{r_{-}^{2}}{2GM}\omega
\end{equation}

\noindent where $L$ can be regarded as modified number which is
connected with angular quantum, global monopole parameter and the
coupling constant. It reduces to $l$ without global monopole in
black hole. Substitute these equations (17), (18), (19) into Eq.
(15) to obtain,

\begin{eqnarray}
y(1-y)\frac{d^{2}F_{\omega l}}{dy^{2}}
+\{(2i\varpi+1)-[(2i\varpi+L+1)+(L+1)+1]y\} \frac{dF_{\omega
l}}{dy}\nonumber\\
-(2i\varpi+L+1)(L+1)F_{\omega l}=0\hspace{3cm}
\end{eqnarray}

\noindent which is the hypergeometric equation [41]. The solution
to Eq. (20) is certainly written as hypergeometric functions [41],

\begin{eqnarray}
F_{\omega l}=C_{1}F(2i\varpi+L+1, L+1, 2i\varpi+1, y)
\hspace{2cm}\nonumber\\
+C_{2}y^{-2i\varpi}F(L+1, L-2i\varpi+1, 1-2i\varpi, y)
\end{eqnarray}

\noindent According to Eq. (17), the radial parts of the solution
is,

\begin{eqnarray}
R_{\omega l}^{n}=C_{1}y^{i\varpi}(1-y)^{L+1}
F(2i\varpi+L+1, L+1, 2i\varpi+1, y)\hspace{2cm}\nonumber\\
+C_{2}y^{-i\varpi}(1-y)^{L+1}F(L+1, L-2i\varpi+1, 1-2i\varpi, y)
\end{eqnarray}

\noindent where $y^{i\varpi}$ term represents an outgoing wave at
the black hole horizon and $y^{-i\varpi}$ term implies an
incoming wave when $y=0$. According to the boundary condition,
only incoming mode exists at near-region , which means $C_{1}=0$.
In the low-frequency limit, the radial part becomes,

\begin{eqnarray}
R_{0l}^{n}=\lim_{\omega\longrightarrow0}R_{\omega
l}^{n}\hspace{3cm}\nonumber\\
=C(1-y)^{L+1}F(L+1, L+1, 1, y)
\end{eqnarray}

\noindent where $C$ is a coefficient. According to the Pfaff
theorem for the hypergeometric function and the Murphy expressions
for the Legendre polynomials [41], the asymptotic form of the
radial part is,

\begin{equation}
R_{0l}^{n}=C(-1)^{L}P_{L}(1-\frac{1-8\pi G\eta^{2}}{GM}r)
\end{equation}

In the region near the outer event horizon where $r-r_{-}\gg MG$,
the mass term $2GM/r$ in the metric (2) vanishes, so the radial
wave equation is changed as

\begin{eqnarray}
x(1-x)\frac{d^{2}R_{\omega l}^{f}}{dx^{2}}+(1-3x)\frac{dR_{\omega
l}^{f}}{dx}\hspace{4cm}\nonumber\\
+[6\xi-\frac{l(l+1)-16\pi\xi G\eta^{2}}{(1-8\pi G\eta^{2})(1-x)}
+\frac{\omega^{2}}{\psi_{0}^{2}}\frac{1-x}{x}]R_{\omega l}^{f}=0
\end{eqnarray}

\noindent where

\begin{equation}
x=1-\frac{\psi_{0}r}{1-8\pi G\eta^{2}}
\end{equation}

\noindent and,

\begin{equation}
R_{\omega l}^{f}=x^{i\tilde{\omega}}(1-x)^{L}\tilde{F}_{\omega
l}
\end{equation}

\begin{equation}
\tilde{\omega}=\frac{\omega}{\psi_{0}}
\end{equation}

\noindent The wave equation is denoted as a hypergeometric
equation,

\begin{eqnarray}
x(1-x)\frac{d^{2}\tilde{F}_{\omega l}}{dx^{2}}
+[(2i\tilde{\omega}+1)-(2i\tilde{\omega}+2L+3)x]\frac{d\tilde{F}_{\omega
l}}{dx}\hspace{1cm}\nonumber\\
-[2i\tilde{\omega}(L+1)+L(L+2)-6\xi]\tilde{F}_{\omega l}=0
\end{eqnarray}

\noindent The solution to Eq. (29) is [41],

\begin{equation}
\tilde{F}_{\omega l}=D_{1}F(\alpha_{+}, \alpha_{-}, \gamma, x)
+D_{2}x^{1-\gamma}F(\alpha_{+}-\gamma+1, \alpha_{-}-\gamma+1,
2-\gamma, x)
\end{equation}

\noindent where

\begin{equation}
\alpha_{\pm}=(i\tilde{\omega}+L+1)
\pm\sqrt{-\tilde{\omega}^{2}-2i\tilde{\omega}-2L-1+6\xi}
\end{equation}

\begin{equation}
\gamma=2i\tilde{\omega}+1
\end{equation}

\noindent The solution near the outer horizon is,

\begin{eqnarray}
R_{\omega l}^{f}=D_{1}x^{i\tilde{\omega}}(1-x)^{L} F(\alpha_{+},
\alpha_{-}, \gamma, x)\hspace{5cm}\nonumber\\
+D_{2}x^{-i\tilde{\omega}}(1-x)^{L}F(\alpha_{+}-\gamma+1,
\alpha_{-}-\gamma+1, 2-\gamma, x)
\end{eqnarray}

\noindent When the radial coordinate approaches the larger radius
of the metric, then $\lim\limits_{r\longrightarrow r_{+}}{f(r)}=1-8\pi
G\eta^{2}-\psi_{0}r$, and we can write the tortoise coordinate as
$e^{i\omega r_{\ast}}=x^{-i\tilde{\omega}}$, which leading the
solution (33) to be

\begin{equation}
R_{\omega l}^{f}\approx D_{1}x^{i\tilde{\omega}}
+D_{2}x^{-i\tilde{\omega}}
\end{equation}

\noindent At this position we combine Eq. (12) with Eq. (10) to
obtain,

\begin{equation}
R_{\omega l}^{f}\approx \frac{A_{\omega
l}^{in}}{r_{+}}x^{i\tilde{\omega}} +\frac{A_{\omega
l}^{out}}{r_{+}}x^{-i\tilde{\omega}}
\end{equation}

\noindent and then,

\begin{equation}
A_{\omega l}^{in}=r_{+}D_{1}
\end{equation}

\begin{equation}
A_{\omega l}^{out}=r_{+}D_{2}
\end{equation}

\noindent It is clear that the greybody factors can be shown in
terms of the coefficients $D_{1}$ and $D_{2}$ from Eq. (11) [27],

\begin{equation}
\gamma_{l}(\omega)=1-|\frac{D_{2}}{D_{1}}|^{2}
\end{equation}

It is reasonable to match the two asymptotic solutions in Eq. (22)
and Eq. (33) in the overlapping region of their own districts to
relate coefficients $D_{1}$ and $D_{2}$ with $C_{1}$ or $C_{2}$
mentioned in Eq. (22) like Ref. [27]. We are also interested in
the near-region solution with larger radial coordinate and the
far-region solution with smaller $r$. Using the $y\to 1-y$
transformation law for the hypergeometric function, Eq. (22) is
expressed as

\begin{eqnarray}
R_{\omega l}^{n}=C_{2}y^{-i\varpi}(\frac{2GM}{1-8\pi G\eta^{2}}
\frac{1}{r})^{L+1}\frac{\Gamma(1-2i\varpi)\Gamma(-2L-1)}
{\Gamma(-2i\varpi-L)\Gamma(-L)}\hspace{0.5cm}\nonumber\\
\times F(L+1, L-2i\varpi+1, 2L+2; 1-y)\nonumber\\
+C_{2}y^{-i\varpi}(\frac{2GM}{1-8\pi G\eta^{2}}
\frac{1}{r})^{-L}\frac{\Gamma(1-2i\varpi)\Gamma(2L+1)}
{\Gamma(L+1)\Gamma(L-2i\varpi+1)}\nonumber\\
\times F(-2i\varpi-L, -L, -2L; 1-y)\hspace{1cm}
\end{eqnarray}

\noindent The hypergeometric functions in Eq. (33) are also
transformed to give rise to,

\begin{eqnarray}
R_{\omega l}^{f}=D_{1}x^{i\tilde{\omega}}(\frac{\psi_{0}r}{1-8\pi
G\eta^{2}})^{L}\frac{\Gamma(\gamma)\Gamma(\gamma-\alpha_{+}-\alpha_{-})}
{\Gamma(\gamma-\alpha_{+})\Gamma(\gamma-\alpha_{-})}\hspace{3cm}\nonumber\\
\times F(\alpha_{+}, \alpha_{-}, \alpha_{+}+\alpha_{-}-\gamma+1;
1-x)\hspace{1cm}\nonumber\\
+D_{1}x^{i\tilde{\omega}}(\frac{\psi_{0}r}{1-8\pi
G\eta^{2}})^{-L-1}\frac{\Gamma(\gamma)\Gamma(\alpha_{+}+\alpha_{-}-\gamma)}
{\Gamma(\alpha_{+})\Gamma(\alpha_{-})}\hspace{2cm}\nonumber\\
\times F(\gamma-\alpha_{+}, \gamma-\alpha_{-},
\gamma-\alpha_{+}-\alpha_{-}+1; 1-x)\nonumber\\
+D_{2}x^{-i\tilde{\omega}}(\frac{\psi_{0}r}{1-8\pi G\eta^{2}})^{L}
\frac{\Gamma(2-\gamma)\Gamma(\gamma-\alpha_{+}-\alpha_{-})}
{\Gamma(1-\alpha_{+})\Gamma(1-\alpha_{-})}\hspace{1.5cm}\nonumber\\
\times F(\alpha_{+}-\gamma+1, \alpha_{-}-\gamma+1,
\alpha_{+}+\alpha_{-}-\gamma+1; 1-x)\nonumber\\
+D_{2}x^{-i\tilde{\omega}}(\frac{\psi_{0}r}{1-8\pi
G\eta^{2}})^{-L-1}\frac{\Gamma(2-\gamma)\Gamma(\alpha_{+}+\alpha_{-}
-\gamma)}{\Gamma(\alpha_{+}-\gamma+1)\Gamma(\alpha_{-}-\gamma+1)}
\nonumber\\
\times F(1-\alpha_{+}, 1-\alpha_{-},
\gamma-\alpha_{+}-\alpha_{-}+1; 1-x)
\end{eqnarray}

\noindent The matching condition is [27],

\begin{equation}
\lim_{y\longrightarrow1}R_{\omega l}^{n}
=\lim_{x\longrightarrow1}R_{\omega l}^{f}
\end{equation}

\noindent We substitute Eq. (39) and Eq. (40) into Eq. (41) and
compare the coefficients of term $r^{L}$ and $r^{-L-1}$
respectively to obtain,

\begin{eqnarray}
D_{1}(\frac{\psi_{0}}{1-8\pi
G\eta^{2}})^{L}\frac{\Gamma(\gamma)\Gamma(\gamma-\alpha_{+}-\alpha_{-})}
{\Gamma(\gamma-\alpha_{+})\Gamma(\gamma-\alpha_{-})}\hspace{2.5cm}\nonumber\\
+D_{2}(\frac{\psi_{0}}{1-8\pi G\eta^{2}})^{L}
\frac{\Gamma(2-\gamma)\Gamma(\gamma-\alpha_{+}-\alpha_{-})}
{\Gamma(1-\alpha_{+})\Gamma(1-\alpha_{-})}\hspace{1cm}\nonumber\\
=C_{2}(\frac{2GM}{1-8\pi G\eta^{2}}
)^{-L}\frac{\Gamma(1-2i\varpi)\Gamma(2L+1)}
{\Gamma(L+1)\Gamma(L-2i\varpi+1)}
\end{eqnarray}

\begin{eqnarray}
D_{1}(\frac{\psi_{0}}{1-8\pi
G\eta^{2}})^{-L-1}\frac{\Gamma(\gamma)\Gamma(\alpha_{+}+\alpha_{-}-\gamma)}
{\Gamma(\alpha_{+})\Gamma(\alpha_{-})}\hspace{2cm}\nonumber\\
+D_{2}(\frac{\psi_{0}}{1-8\pi
G\eta^{2}})^{-L-1}\frac{\Gamma(2-\gamma)\Gamma(\alpha_{+}+\alpha_{-}
-\gamma)}{\Gamma(\alpha_{+}-\gamma+1)\Gamma(\alpha_{-}-\gamma+1)}
\nonumber\\
=C_{2}(\frac{2GM}{1-8\pi G\eta^{2}}
)^{L+1}\frac{\Gamma(1-2i\varpi)\Gamma(-2L-1)}
{\Gamma(-2i\varpi-L)\Gamma(-L)}
\end{eqnarray}

\noindent Having solved Eq. (42) and Eq. (43), we arrive at the
analytic approximation for greybody factor,

\begin{eqnarray}
\gamma_{l}(\omega)=1-\hspace{11.5cm}\nonumber\\
|[(\frac{2\psi_{0}GM}{(1-8\pi G\eta^{2})^{2}})^{2L+1}
\frac{\Gamma(2i\tilde{\omega}+1)\Gamma(1-2i\varpi)
(\Gamma(-2L-1))^{2}}{\Gamma(\gamma-\alpha_{+})
\Gamma(\gamma-\alpha_{-})\Gamma(-L-2i\varpi)\Gamma(-L)}
\hspace{1.5cm}\nonumber\\
-\frac{\Gamma(2i\tilde{\omega}+1)\Gamma(1-2i\varpi)(\Gamma(2L+1))^{2}}
{\Gamma(L-2i\varpi+1)\Gamma(\alpha_{+})\Gamma(\alpha_{-})
\Gamma(L+1)}]\hspace{4cm}\nonumber\\
\times[\frac{\Gamma(1-2i\varpi)\Gamma(1-2i\tilde{\omega})
(\Gamma(2L+1))^{2}}{\Gamma(L-2i\varpi+1)\Gamma(\alpha_{+}-\gamma+1)
\Gamma(\alpha_{-}-\gamma+1)\Gamma(L+1)}\hspace{2.5cm}\nonumber\\
-(\frac{2\psi_{0}GM}{(1-8\pi G\eta^{2})^{2}})^{2L+1}
\frac{\Gamma(1-2i\tilde{\omega})\Gamma(1-2i\varpi)(\Gamma(-2L-1))^{2}}
{\Gamma(1-\alpha_{+})\Gamma(1-\alpha_{-})\Gamma(-L-2i\varpi)\Gamma(-L)}
]^{-1}|^{2}
\end{eqnarray}

\noindent The dependence of greybody factor on the frequency due
to the model parameter $\eta$ is plotted in the Figure groups
consisting of Fig. 1, Fig. 2 and Fig. 3 with angular quantum
number $l=0, 1, 2$ respectively. When the quantum number vanishes,
the greybody factor becomes smaller as the parameter $\eta$
changes to be small. In the case of nonvanishing angular quantum
number, the curves of absorption probability rise while the global
monopole parameter decreases. Noticing the magnitude of greybody factor is getting
smaller as $l$ increases, we extend the frequency axis to 0.01 to
see the behavior of the function.
Combined with the energy emission rate in section IV, we find out
a significant property that with a large decreasing global
monopole parameter the summation of nonvanishing angular quantum
number leads a increasing energy emission, however, the parameter
$\eta$ decreases tiny enough so that the first order greybody
factor $\gamma_{0}(\omega)$ dominates over all others.

In order to understand how the deviation of standard general
relativity effect the absorption probability, we derive the
greybody factor in Schwarzschild spacetime with a global monopole.
Fortunately, the Eq. (25) in the region close to the inner horizon
and the solution belonging to it are suitable for the this new
case with no $f(R)$ gravity involved. Hence, we just need to
resolve Eq. (13) to find the asymptotic solution with regarded
metric $f(r)=1-8\pi G\eta^{2}-\frac{2GM}{r}$. With asymptotic
approach and series expansion at $r\to \infty$ to first order for
each term, Eq. (13) can be written as [42-44],

\begin{equation}
\frac{d^{2}R_{\omega l}^{fM}}{dr^{2}}+\frac{2}{r}
\frac{dR_{\omega l}^{fM}}{dr}+(\frac{\omega^{2}}
{(1-8\pi G \eta^{2})^{2}}-\frac{L(L+1)}{r^{2}})R_{\omega l}^{fM}=0
\end{equation}

\noindent This is a Bessel equation of which the solution in
asymptotic region is given by

\begin{equation}
R_{\omega l}^{fM}=\frac{1}{\sqrt{r}}\left[ B_{1}J_{L+\frac{1}{2}}
(\frac{\omega r}{1-8\pi G\eta^{2}})+B_{2}Y_{L+\frac{1}{2}}
(\frac{\omega r}{1-8\pi G\eta^{2}})\right]
\end{equation}

\noindent Taking $r\to 0$, $R_{\omega l}^{fM}$ at
intermediate-region with low energy can be expressed as

\begin{equation}
R_{\omega l}^{fM}\approx \frac{B_{1}r^{L}}{\Gamma(L+\frac{3}{2})}
\left(\frac{\omega r}{2(1-8\pi G\eta^{2})}\right)^{L+\frac{1}{2}}
-\frac{B_{2}\Gamma(L+\frac{1}{2})}{\pi r^{L+1}}\left(\frac{\omega r}
{2(1-8\pi G\eta^{2})}\right)^{-L-\frac{1}{2}}
\end{equation}

\noindent Compare Eq. (47) with Eq. (39) at $y\to 1$, one may find
the ratio of $B_{1}$ and $B_{2}$

\begin{eqnarray}
B\equiv \frac{B_{1}}{B_{2}}\hspace{13cm}\nonumber\\
=-\frac{1}{\pi}\left(\frac{(1-8\pi G\eta^{2})^{2}}
{GM\omega}\right)^{2L+1}\frac{(L+\frac{1}{2})\Gamma^{2}
(L+\frac{1}{2})\Gamma(2L+1)\Gamma(-L)\Gamma(-L-i\frac{4GM\omega}
{(1-8\pi G \eta^{2})^{2}})}{\Gamma(L+1)\Gamma(L+1-i\frac{4GM\omega}
{(1-8\pi G \eta^{2})^{2}})\Gamma(-2L-1)}
\end{eqnarray}

The absorption probability with low-frequency limit can be
approximated as (see [27]),

\begin{equation}
\gamma_{l}(\omega)\approx\frac{2i(B^{*}-B)}{|B|^{2}}
\end{equation}

\noindent where the star index stands conjugation. We plot the
dependence of the absorption probability on the $f(R)$ gravity
factor $\psi_{0}$ in Fig. 4, 5, 6 as the angular quantum number
$l=0,1,2$ respectively. The solid curves in three graphs show the
behavior of greybody factor of emission without the effect from
gravity correction. For low-frequency ($\omega\ll 1/MG, MG=1$), the
absorption probability curve rises as the deviation from the general
relativity $\psi_{0}$ increases, which means the more deviation
exists, the larger greybody factor of a emitted scalar field
coupling to gravitational field can be obtained. The total order
of magnitude of greybody factor gets smaller as the angular index
increases. This feature conform with the absorption probability
$\gamma_{l}\approx \omega^{2l+2}$ in low-frequency approximation
$\omega\ll 1/MG$, which shows the first partial absorption
probability is the leadership. It should be
pointed out that the greybody factor is an increasing function of
frequency no matter how many the angular quantum number is equal
to and how great the model parameter or the deviation from general
relativity is.

\vspace{0.8cm} \noindent \textbf{IV.\hspace{0.4cm}The energy
emission rate and the generalized absorption cross section
with an $f(R)$ global monopole}

Here we discuss the flux spectrum  which is the number of massless
scalar particles emitted by the gravitational source per unit time
and is given by [27, 39],

\begin{equation}
\frac{dN(\omega)}{dt}=\frac{1}{2\pi}\frac{1}{e^{\frac{\omega}{T_{H}}}-1}
(\sum_{l=0}^{\infty}(2l+1)\gamma_{l}(\omega))d\omega
\end{equation}

\noindent leading the differential energy emission rate [27, 39],

\begin{eqnarray}
\frac{d^{2}E(\omega)}{dtd\omega}=\frac{d^{2}N(\omega)}{dtd\omega}
\omega\hspace{1.5cm}\nonumber\\
=\frac{1}{2\pi}\frac{\omega}{e^{\frac{\omega}{T_{H}}}-1}
\sum_{l=0}^{\infty}(2l+1)\gamma_{l}(\omega)
\end{eqnarray}

\noindent where $T_{H}$ is the Hawking temperature which is
$\frac{1}{4\pi}\left(\frac{1-8\pi
G\eta^{2}}{r_{-}}-2\psi_{0}\right)$. We show our numerical results
for the differential energy emission rate in Fig. 7. One can see
as the model parameter $\eta$ reduces, the maximum of emission
rate increases. While $\eta$ decreases to some value fitting a
typical grand unified theory, the curve of energy emission rate
declines. This phenomenon presents that a stronger global monopole
parameter alter the attribute of emission of a scalar field
coupling to the gravitational field. In Fig.8 the influence from
$f(R)$ theory on the dependence of the differential energy
emission rate on the frequency is described. It is evident that
the greater deviation from the standard general relativity leads
the curves to drop while moves left. The curves with larger
$\psi_{0}$ keep more complete feature of emission rate
distinctively for low frequency $\omega \ll 1/MG$ and $MG=1$. The
shapes of all of the curves subject to the global monopole
parameter and the corrections to the general relativity are
similar.

We start to investigate the generalized absorption cross section
of a scalar field coupling to the gravitational field in the
background of a Schwarzschild black hole with a $f(R)$ global
monopole. The absorption cross section for the emission of a
particles from a black hole can be defined as
$\sigma\equiv\frac{absorbed \ flux}{incident \ wave \ current}$
leading the expression for asymptotically flat spherically
symmetric spacetime as [27],

\begin{eqnarray}
\sigma=\sum_{l=0}^{\infty}\sigma_{l}
=\frac{\pi}{\omega^{2}}\sum_{l=0}^{\infty}(2l+1)\gamma_{l}(\omega)
\end{eqnarray}

\noindent where $\sigma_{l}$ stands for the absorption cross
section of each partial wave. We continue to discuss the condition
for applying Eq. (52). We follow Ref. [27] to choose the flux
corresponding to the scatted wave and incident current

\begin{equation}
F=\frac{\pi}{\omega}\sum_{l=0}^{\infty}(2l+1)\gamma_{l}(\omega)
\end{equation}

\begin{equation}
|\overrightarrow{J}_{inc}|=\omega \frac{r_{*}^{2}}{r^{2}}
\sqrt{\frac{1}{f}cos^{2}\theta+\frac{r_{*}^{2}}{r^{2}}sin^{2}\theta}
\end{equation}

\noindent Once the incident current equals to $\omega$, Eq. (52)
will be confirmed. In matching intermediate region where $M\ll
r-r_{-}\ll 1/\omega$, there exist $f(r)\approx 1-8\pi G\eta^{2}$
and $r_{*}\approx \frac{r}{1-8\pi G\eta^{2}}$ for a Schwarzschild
spacetime with a f(R) global monopole. If $8\pi G\eta^{2}\ll 1$,
then $f(r)\approx 1$ and $r_{*}\approx r$ which yield
$|\overrightarrow{J}_{inc}|=\omega$. For this reason, the
absorption cross section expression (52) is applicable to the
emission in this background only when $8\pi G\eta^{2}\ll 1$ and
$\psi_{0}\sim 0$ for intermediate region. According to the
greybody factor (44), we also construct the generalized absorption
cross section as a function of frequency in Fig. 9 and Fig. 11.
Both the global model parameter $\eta$ and deviation of general
relativity $\psi_{0}$ will modify the cross section. It is obvious
that almost all curves with different values of $\eta$ or
$\psi_{0}$ boil down together while they separate distinctly in
low frequency, which conform to the low-energy approximation.
From Fig. 9 and the log plot Fig. 10, the larger the cross section
in low-frequency becomes as the larger global monopole parameter.
Moreover, the curves of the general absorption cross section rise
when the variable $\psi_{0}$ increases. We notice that while
$\omega\to 0$, all the values of $\sigma(0)$ for different
$\psi_{0}$ and $\eta$ are finite in Fig. 10 and Fig. 12, and
$\omega\to 0$ reduces as $\psi_{0}$ and $\eta$. It should be
pointed out that the shapes of these curves of the generalized
absorption cross section associated with the frequency are similar
although the cross section is controlled by the global monopole
parameter $\eta$ and variable $\psi_{0}$ in the $f(R)$ theory.

\vspace{0.8cm} \noindent \textbf{VI.\hspace{0.4cm}Discussion and
conclusion}

We discuss the greybody factor for massless scalar field in the
spacetime of gravitational source involving a global monopole in
the context of $f(R)$ gravity theory and we also further study the
energy emission rate and generalized absorption cross section.
These results all exhibit the effect of global monopole and the
influence from $f(R)$ approach. Having matched the two asymptotic
solutions to the field equation at the inner and outer horizons
respectively, we obtain the acceptable expression of the greybody
factor. Further the energy emission rate and the generalized
absorption cross section are also found. It is interesting that
the own effects of global monopole model parameter and the
variable describing the deviation from general relativity appear
in the greybody factor, energy emission rate and the generalized
absorption cross section. The greybody factor becomes smaller due
to the smaller global monopole parameter with $l=0$, which is just
the contrary with novanishing angular quantum number. The larger
the deviation grows, the higher the greybody factor curve rises.
Moreover, a distinct feature of the energy emission rate is the
curve rises as $\eta$ reduces when $8\pi G\eta^{2}\ll 1$ for the
intermediate condition and $8\pi G\eta^{2}\approx 10^{-5}$ for a
typical unified theory. The emission rate curves fall when $\eta$
continues decreasing. For $\psi_{0}\ll 1$, the greater $\psi_{0}$
leads smaller peak values of emission rate in low-energy.  The
generalized absorption cross section becomes larger with respect
to the increasing variable $\eta$ belonging to the global monopole
or increasing $\psi_{0}$ to $f(R)$ gravity. As frequency
$\omega\to 0$, the values of cross section are finite and drop as
$\eta$ and $\psi_{0}$. The cross sections will also be adjusted by
the variable $\psi_{0}$ as the description of modified general
relativity, but their curves keep oscillating around the
frequency. The related topics need to be studied in future.

\vspace{1cm}
\noindent \textbf{Acknowledge}

This work is supported by NSFC No. 10875043 and is partly
supported by the Shanghai Research Foundation No. 07dz22020.

\newpage
\begin{figure}
\setlength{\belowcaptionskip}{10pt} \centering
\includegraphics[width=15cm]{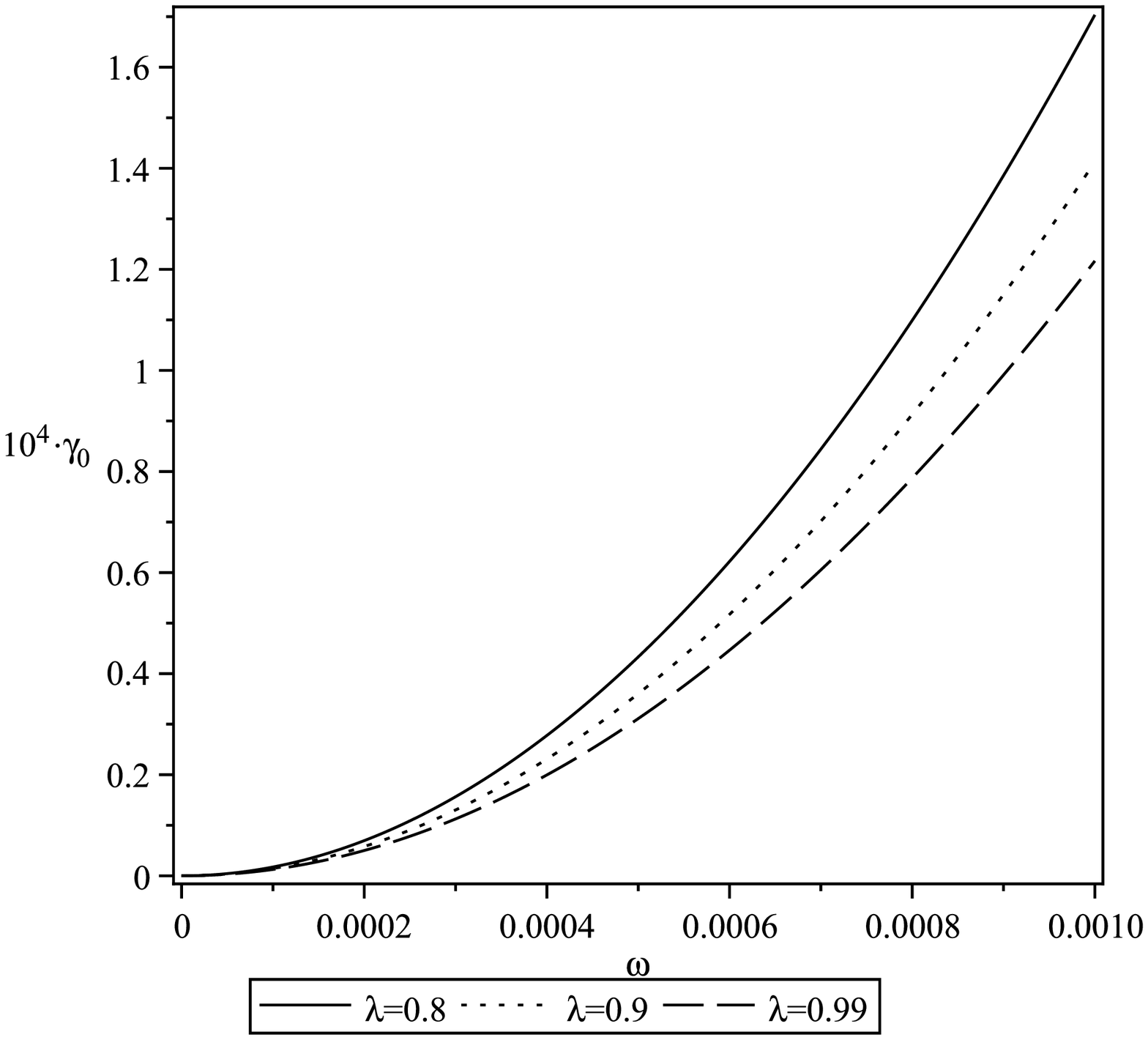}
\caption{The curves of greybody factor with angular quantum number
$l=0$ as a function of the frequency for $\lambda=0.8, 0.9,0.99$
respectively and $\lambda=1-8\pi G\eta^{2}$, $GM=10$, $\psi_{0}=0.02$, $\xi=\frac{1}{12}$.}
\end{figure}

\newpage
\begin{figure}
\setlength{\belowcaptionskip}{10pt} \centering
  \includegraphics[width=15cm]{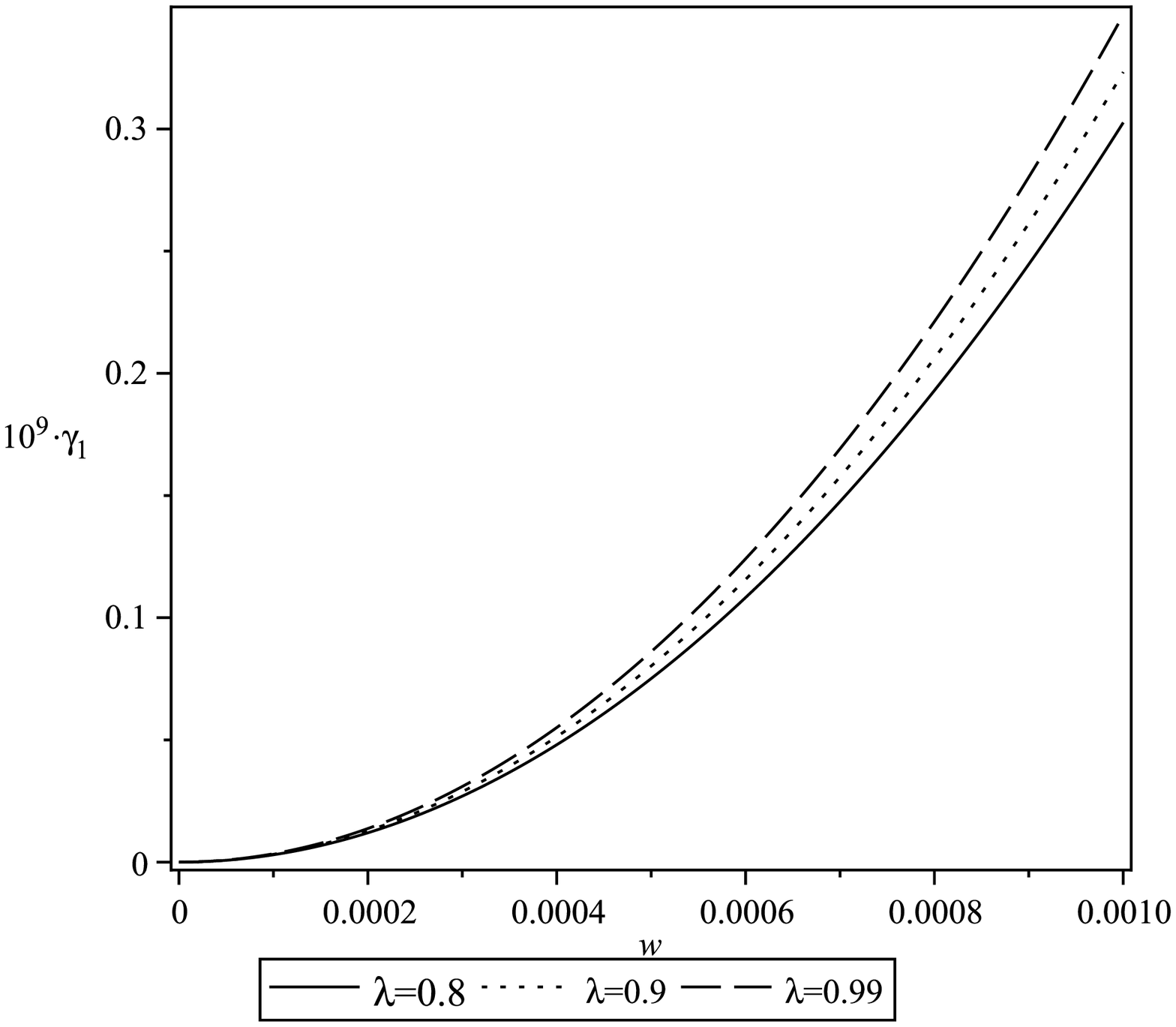}
  \caption{The curves of greybody factor with angular quantum number
$l=1$ as a function of the frequency for $\lambda=0.8, 0.9,0.99$
respectively and $\lambda=1-8\pi G\eta^{2}$, $GM=10$, $\psi_{0}=0.02$, $\xi=\frac{1}{12}$.}
\end{figure}

\newpage
\begin{figure}
\setlength{\belowcaptionskip}{10pt} \centering
  \includegraphics[width=15cm]{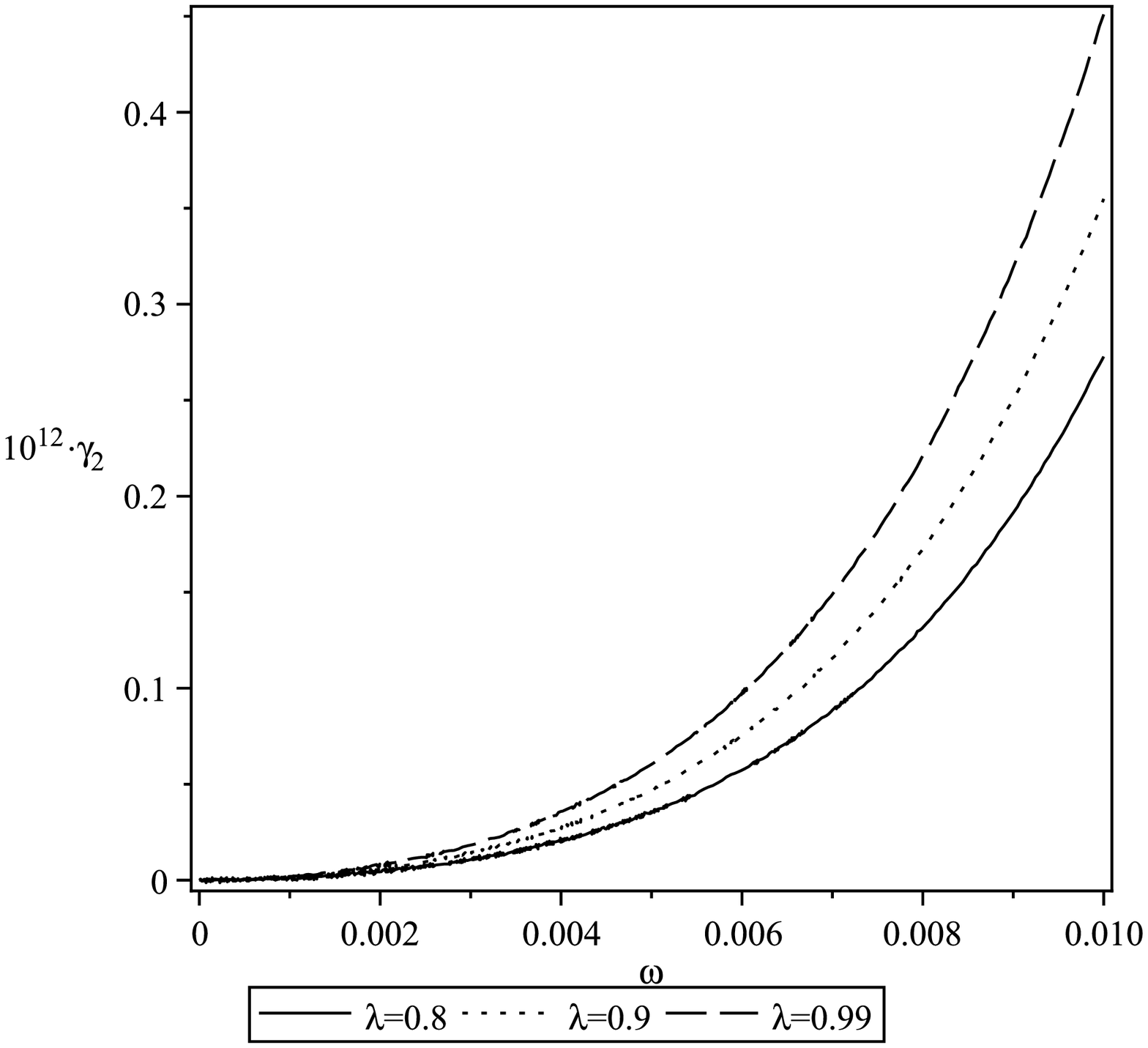}
  \caption{The curves of greybody factor with angular quantum number
$l=2$ as a function of the frequency for $\lambda=0.8, 0.9,0.99$
respectively and $\lambda=1-8\pi G\eta^{2}$, $GM=10$, $\psi_{0}=0.02$, $\xi=\frac{1}{12}$.}
\end{figure}

\newpage
\begin{figure}
\setlength{\belowcaptionskip}{10pt} \centering
  \includegraphics[width=15cm]{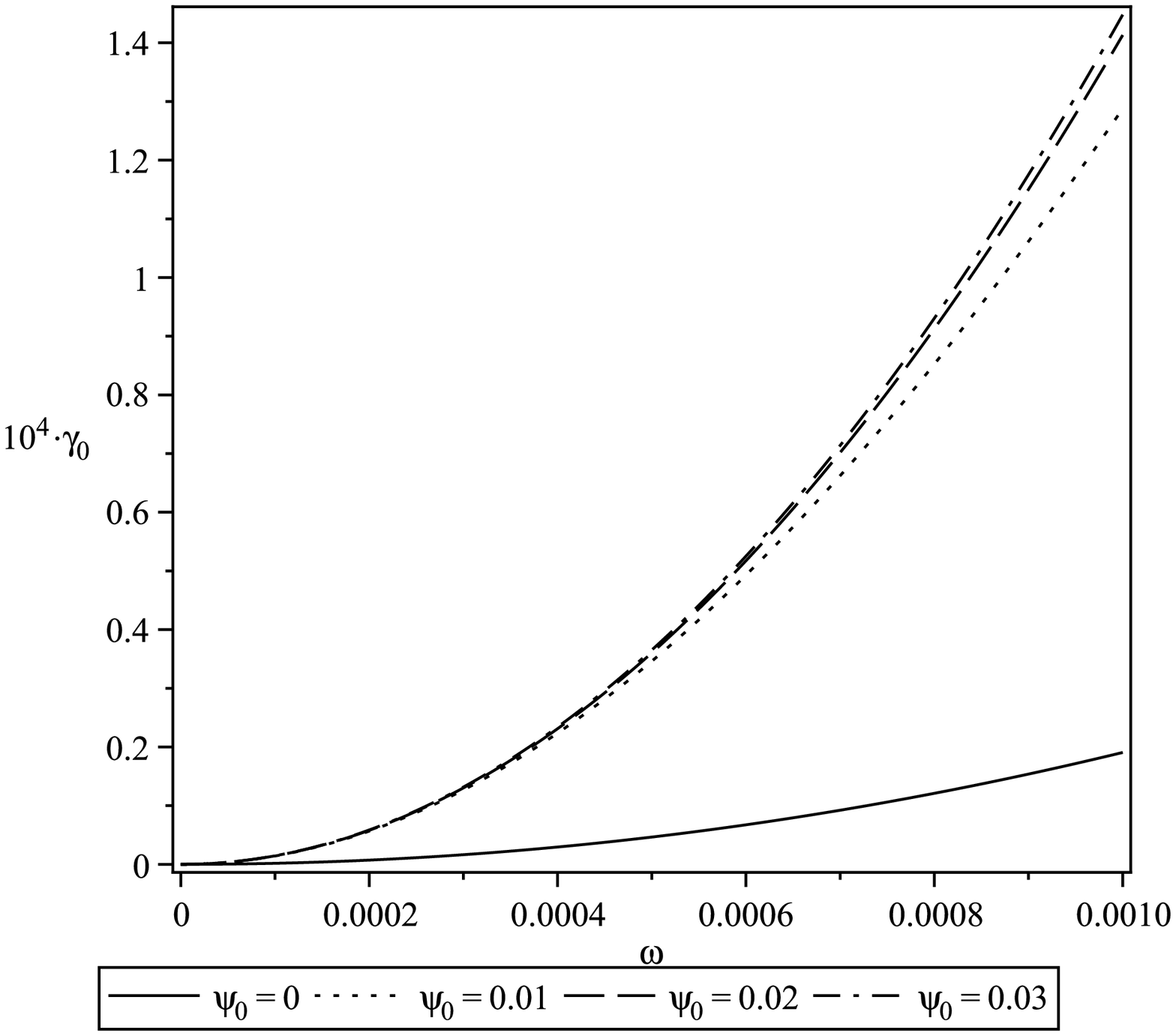}
  \caption{The dependence of Greybody factor for a scalar emission with angular quantum number
$l=0$ on $f(R)$ gravity factor
$\psi_{0}=0,0.01, 0.02, 0.03$ respectively for $GM=1$, $\lambda=0.9$, $\xi=\frac{1}{12}$.}
\end{figure}

\newpage
\begin{figure}
\setlength{\belowcaptionskip}{10pt} \centering
  \includegraphics[width=15cm]{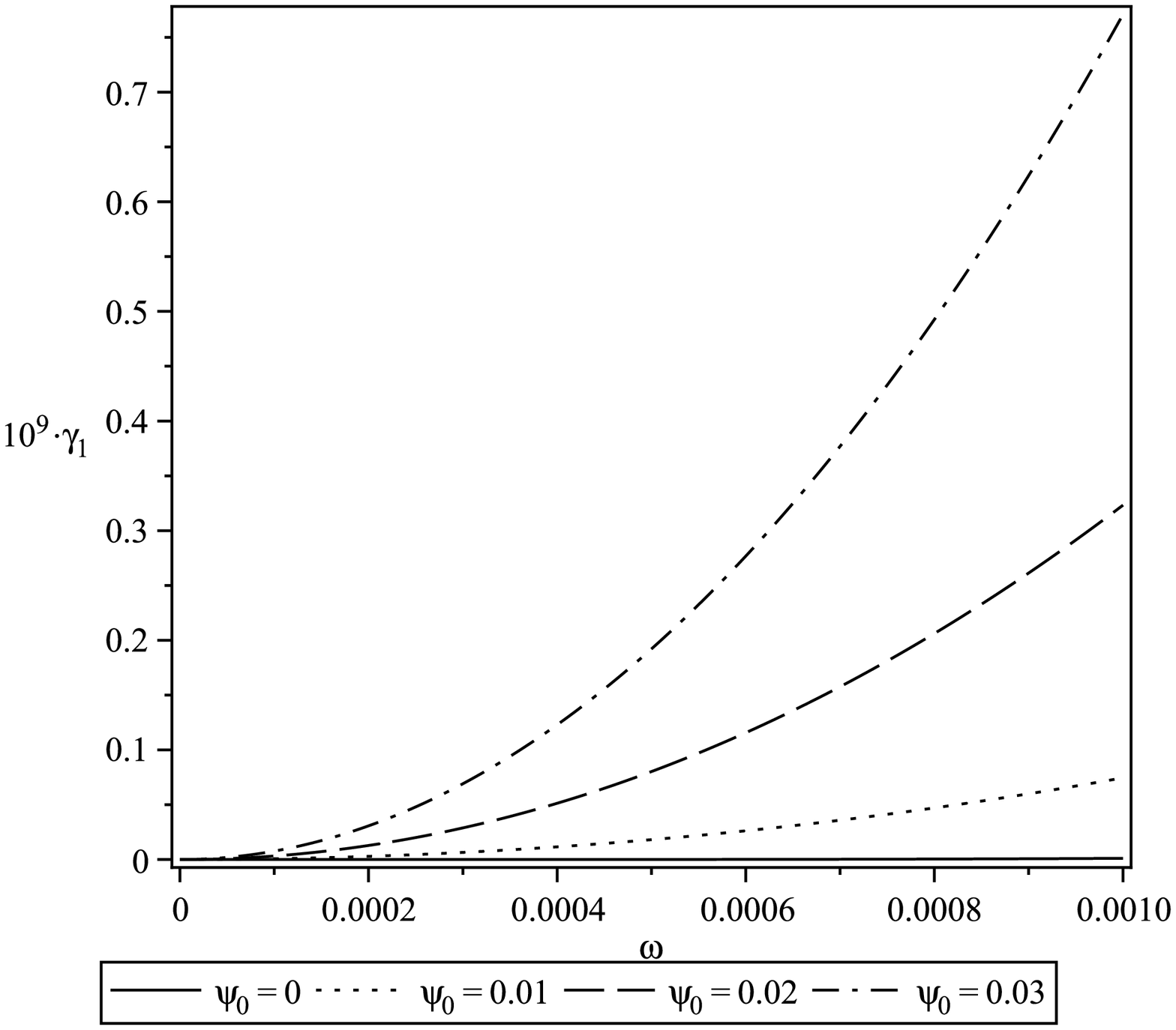}
  \caption{The dependence of Greybody factor for a scalar emission with angular quantum number
$l=1$ on $f(R)$ gravity factor
$\psi_{0}=0,0.01,0.02, 0.03$ respectively for $GM=1$, $\lambda=0.9$, $\xi=\frac{1}{12}$.}
\end{figure}

\newpage
\begin{figure}
\setlength{\belowcaptionskip}{10pt} \centering
  \includegraphics[width=15cm]{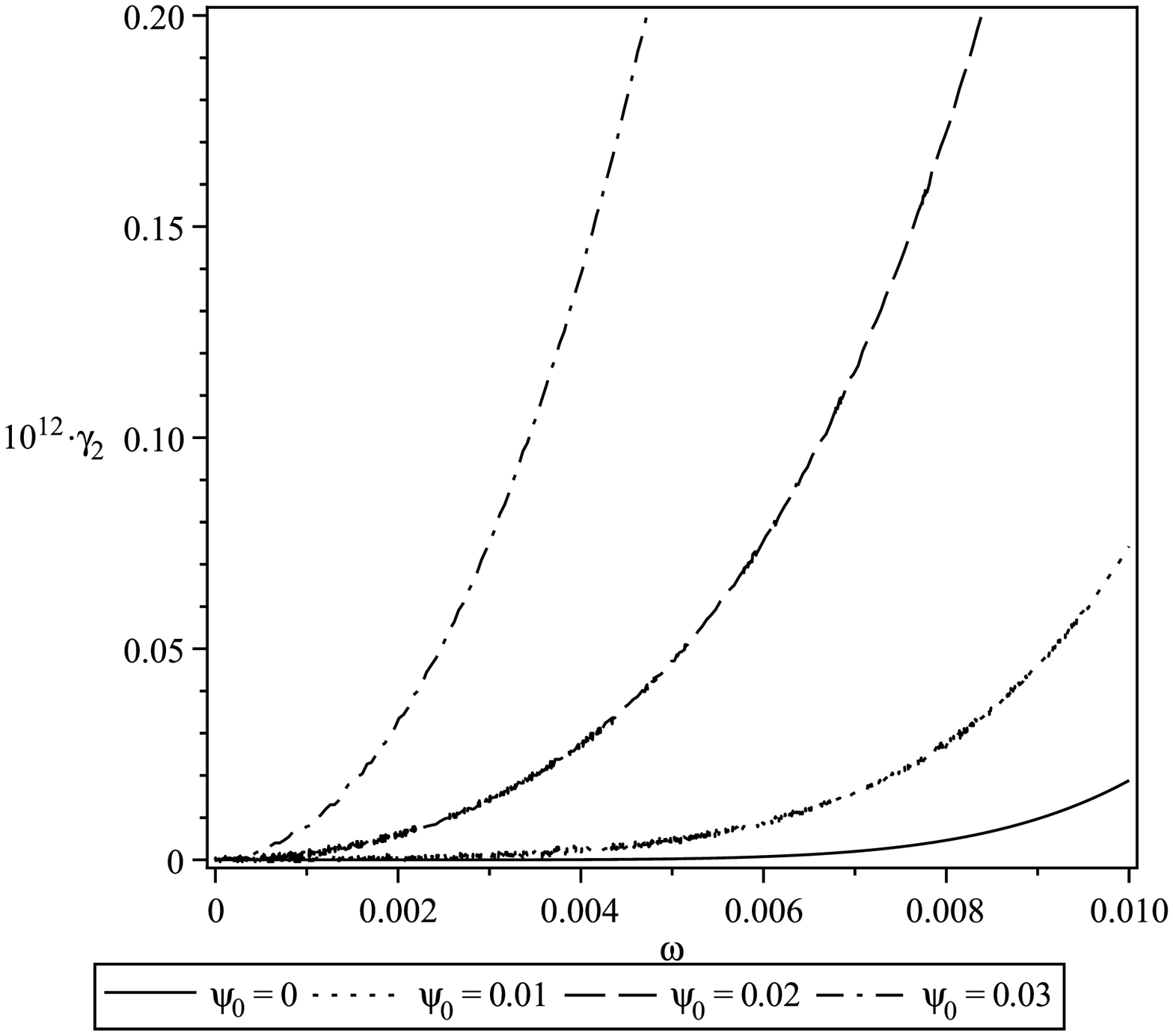}
  \caption{The dependence of Greybody factor for a scalar emission with angular quantum number
$l=2$ on $f(R)$ gravity factor
$\psi_{0}=0,0.01,0.02, 0.03$ respectively for $GM=1$, $\lambda=0.9$, $\xi=\frac{1}{12}$.}
\end{figure}

\newpage
\begin{figure}
\setlength{\belowcaptionskip}{10pt} \centering
  \includegraphics[width=15cm]{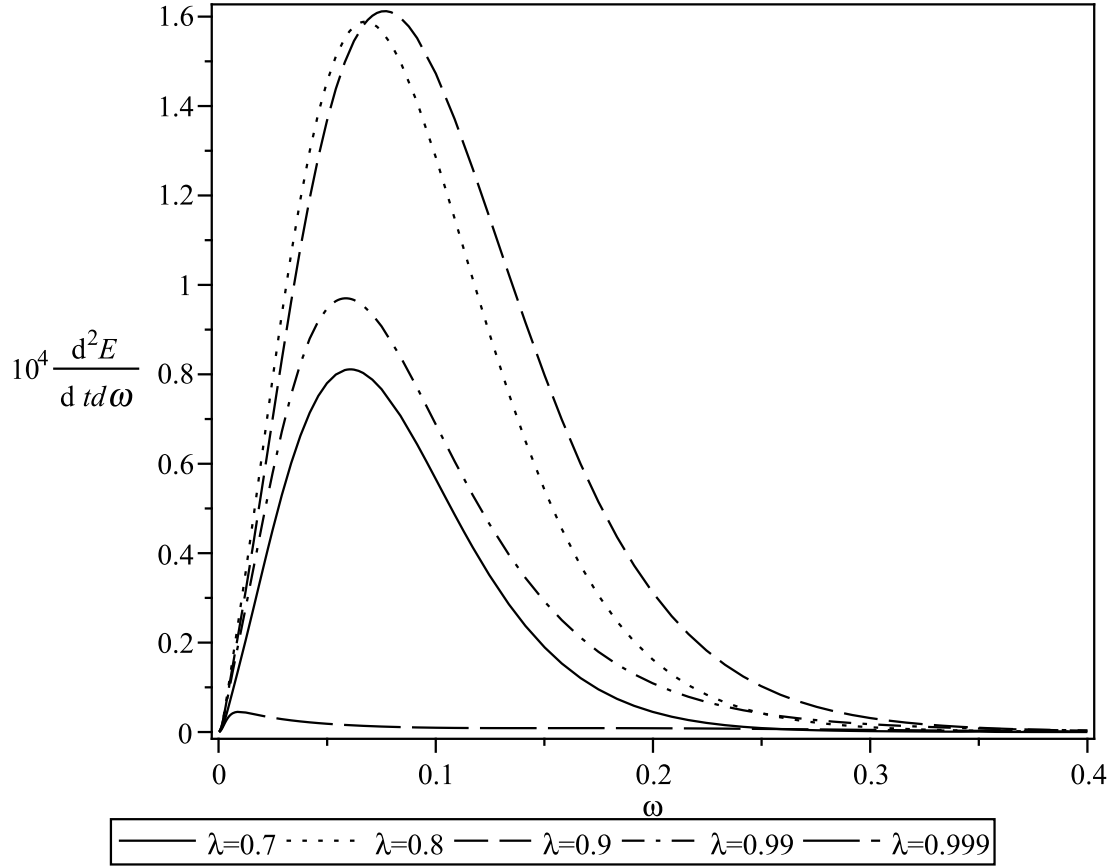}
  \caption{The differential energy emission rate for a scalar emission
   for various $\lambda=0.7, 0.8, 0.9, 0.99, 0.999$ and $\lambda=1-8\pi
   G\eta^{2}$. Here $GM=1$, $\psi_{0}=0.02$, $\xi=\frac{1}{12}$}
\end{figure}

\newpage
\begin{figure}
\setlength{\belowcaptionskip}{10pt} \centering
  \includegraphics[width=15cm]{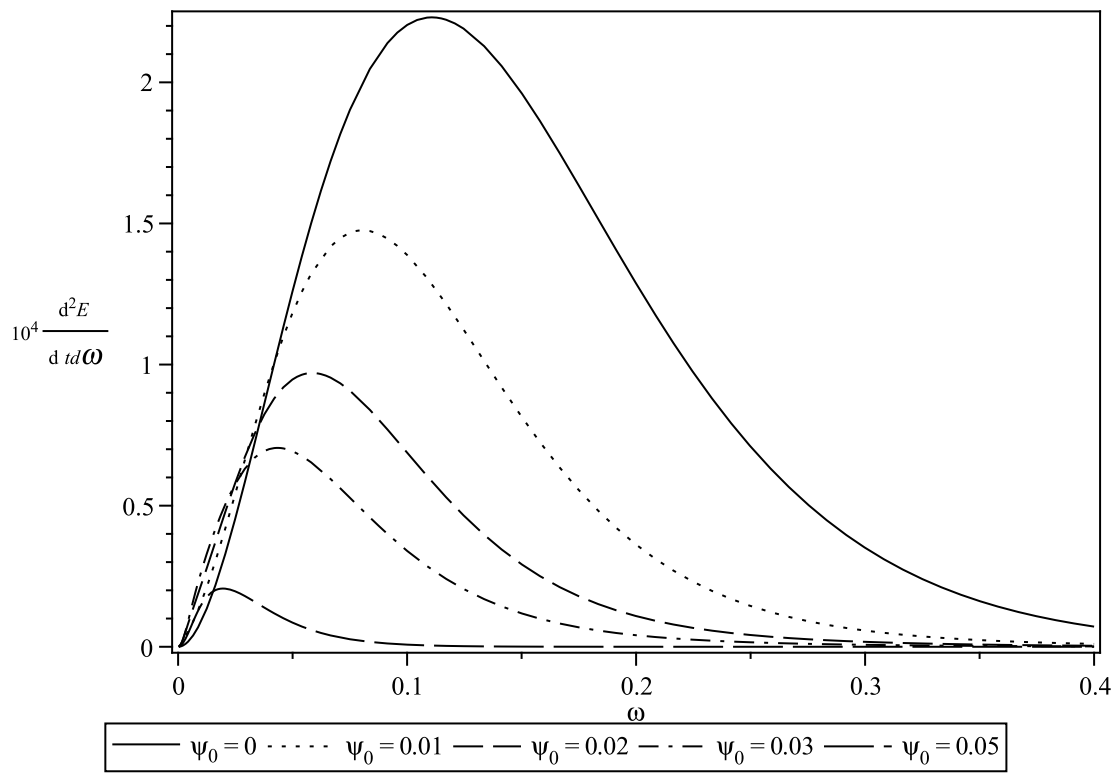}
  \caption{The differential energy emission rate for a scalar emission
   for different $f(R)$ gravity factor
$\psi_{0}=0, 0.01, 0.02, 0.03, 0.05$ respectively. Here $GM=1$, $\lambda=0.99$, $\xi=\frac{1}{12}$}
\end{figure}

\begin{figure}
\setlength{\belowcaptionskip}{10pt} \centering
  \includegraphics[width=15cm]{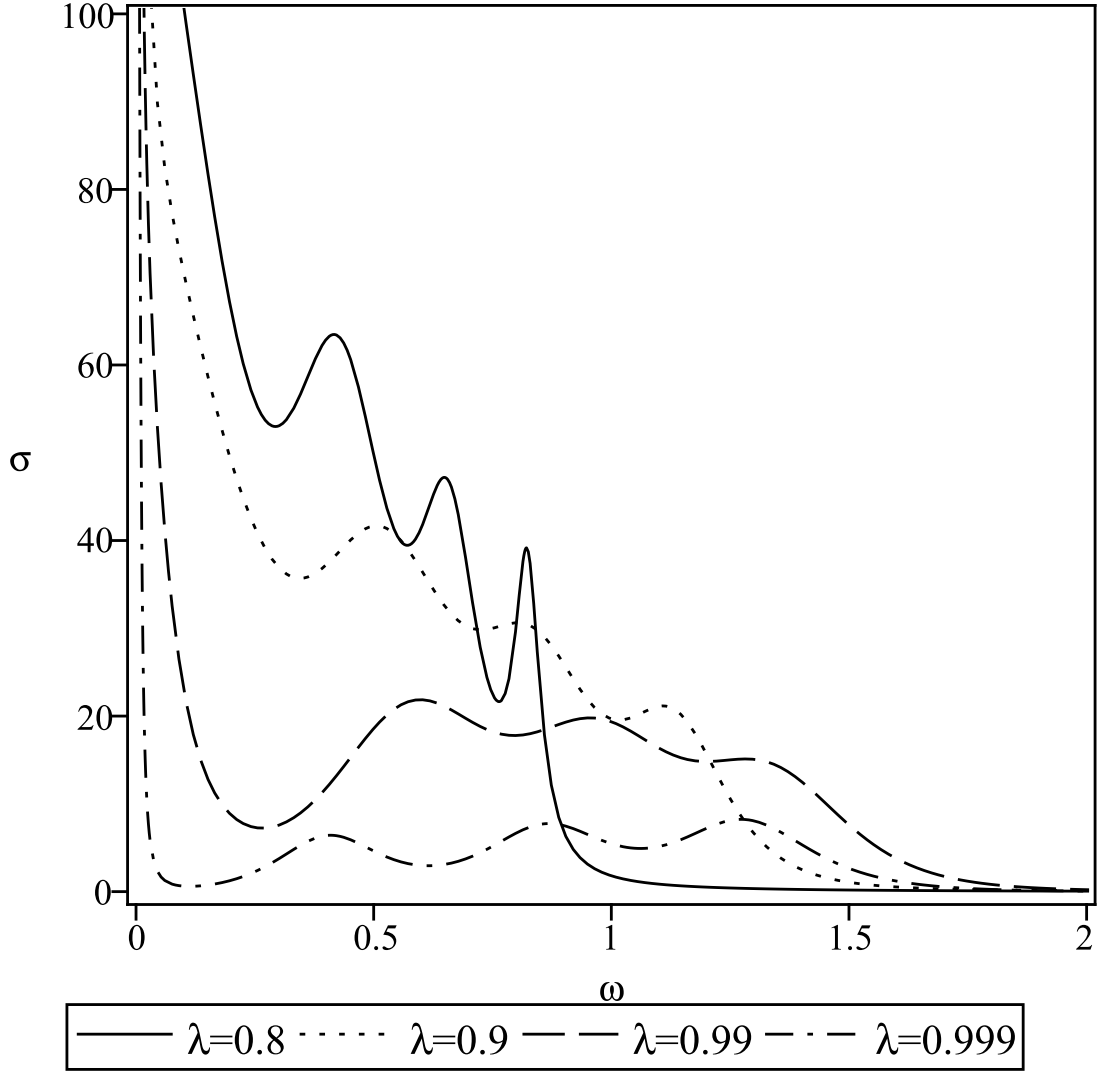}
  \caption{The behavor of the generalized absorption cross section
  as a function of the frequency for $\lambda=0.8, 0.9, 0.99, 0.999$
respectively and $\lambda=1-8\pi G\eta^{2}$. Here $GM=1$, $\psi_{0}=0.02$, $\xi=\frac{1}{12}$.}
\end{figure}

\newpage
\begin{figure}
\setlength{\belowcaptionskip}{10pt} \centering
  \includegraphics[width=15cm]{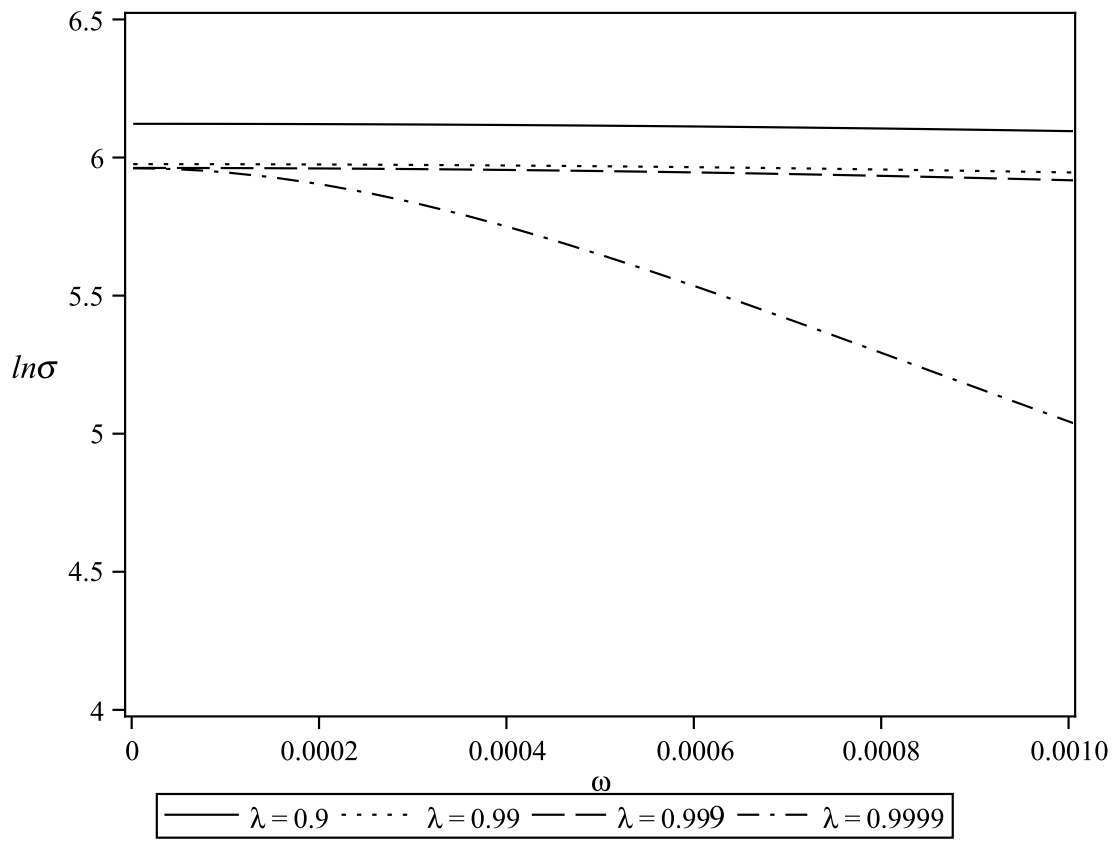}
  \caption{The logarithmic generalized absorption cross section
   depends on global monopole parameter
$\lambda=0.8, 0.9, 0.99, 0.999$ respectively. Here $GM=1$, $\psi_{0}=0.02$, $\xi=\frac{1}{12}$.}
\end{figure}

\newpage
\begin{figure}
\setlength{\belowcaptionskip}{10pt} \centering
  \includegraphics[width=15cm]{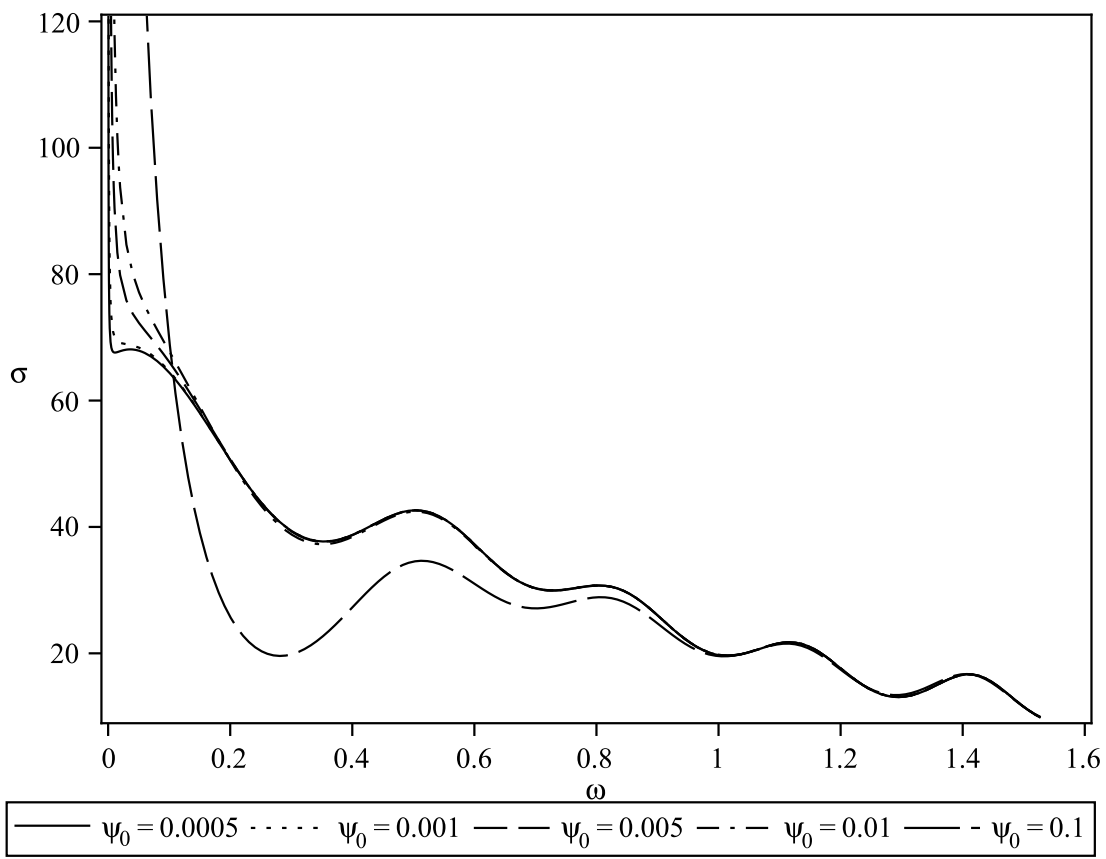}
  \caption{The behavior of the generalized absorption cross section
  as a function of the frequency for $f(R)$ gravity factor
$\psi_{0}=0.0005, 0.001, 0.005, 0.01, 0.1$ respectively.
$GM=1$, $\lambda=0.9$, $\xi=\frac{1}{12}$.}
\end{figure}

\newpage
\begin{figure}
\setlength{\belowcaptionskip}{10pt} \centering
  \includegraphics[width=15cm]{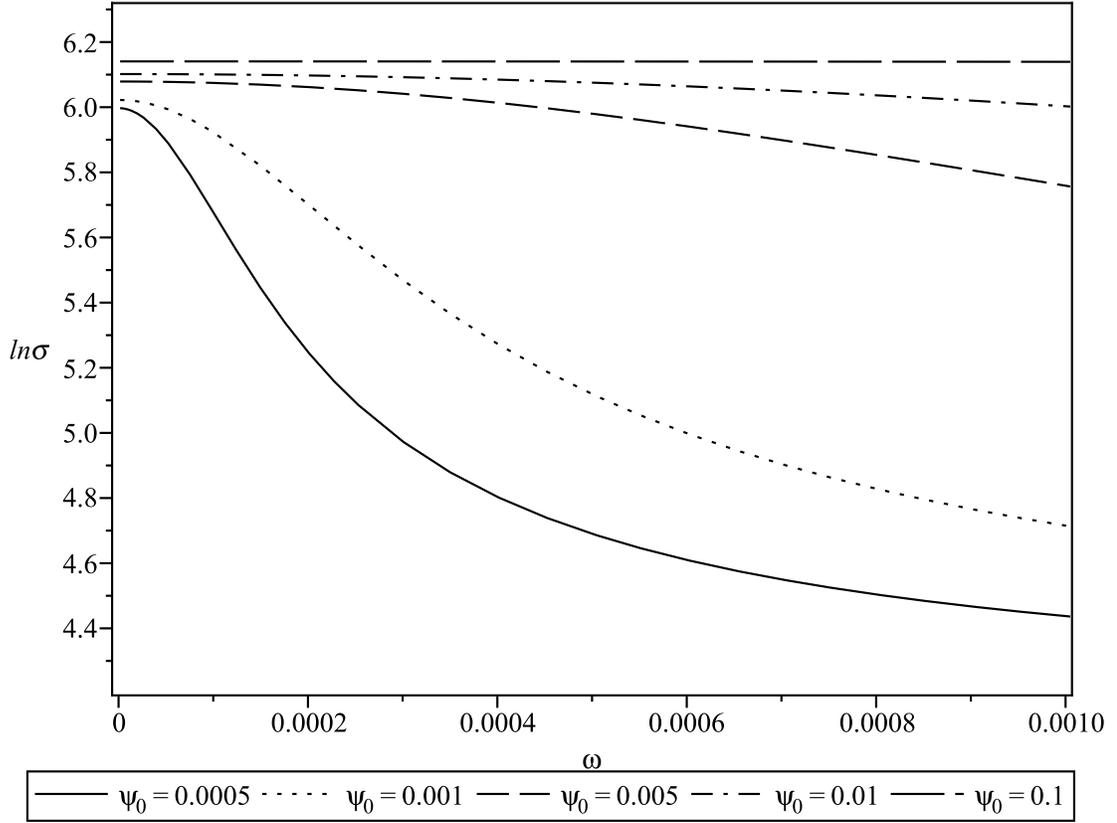}
  \caption{The logarithmic generalized absorption cross section
     depends on $f(R)$ gravity factor
$\psi_{0}=0.0005, 0.001, 0.005, 0.01, 0.1$ respectively.$GM=1$, $\lambda=0.9$, $\xi=\frac{1}{12}$.}
\end{figure}

\end{document}